# ORMIR_XCT: A Python package for high resolution peripheral quantitative computed tomography image processing

Michael T. Kuczynski[1,2], Nathan J. Neeteson[1,2], Kathryn S. Stok[3], Andrew J. Burghardt[4], Michelle A. Espinosa Hernandez[3,5,6], Jared Vicory[7], Justin J. Tse[1,8], Pholpat Durongbhan[3], Serena Bonaretti[9], Andy Kin On Wong[6,10], Steven K. Boyd[1,8], Sarah L. Manske[1,8]

[1]McCaig Institute for Bone and Joint Health, University of Calgary, Calgary, Canada
[2]Biomedical Engineering Graduate Program, University of Calgary, Calgary, Canada
[3]Department of Biomedical Engineering, The University of Melbourne, Parkville, Australia
[4]Department of Radiology and Biomedical Imaging, University of California, San Francisco, USA
[5]Rehabilitation Sciences Institute, The University of Toronto, Toronto, Canada
[6]Joint Department of Medical Imaging, University Health Network, Toronto, Canada
[7]Kitware, Inc., Carrboro, North Carolina, USA
[8]Department of Radiology, Cumming School of Medicine, University of Calgary, Calgary, Canada
[9]Swiss Center for Musculoskeletal Imaging, Balgrist Campus, Zurich, Switzerland
[10]Department of Epidemiology, Dalla Lana School of Public Health, University of Toronto, Toronto, Canada

# 1 – Statement of Need:

High resolution peripheral quantitative computed tomography (HR-pQCT) is an imaging technique with a nominal isotropic voxel size of 61μm capable of imaging trabecular bone *in-vivo*. HR-pQCT has a wide range of applications, primarily focused on bone to improve our understanding of musculoskeletal diseases [1], assess epidemiological associations [2], and evaluate the effects of pharmaceutical interventions [3]. Processing HR-pQCT images has largely been supported using the scanner manufacturer's scripting language (Image Processing Language, IPL, Scanco Medical). However, by expanding image processing workflows outside of the scanner manufacturer's software environment, users have the flexibility to apply more advanced mathematical techniques and leverage modern software packages to improve image processing. The ORMIR_XCT Python package was developed to reimplement some existing IPL workflows and provide an open and reproducible package allowing for the development of advanced HR-pQCT data processing workflows.

# 2 – Package Summary:

The development of this package began during the Jupyter Community Workshop in Maastricht, Netherlands in June 2022, hosted by the Open and Reproducible Musculoskeletal Imaging Research (ORMIR) group. During this workshop, the conceptualization and initial development of the ORMIR_XCT package began with support from Kitware. The ORMIR_XCT package is currently being used by members of the ORMIR community to assess bone mineral density (BMD) and joint space width (JSW) changes in HR-pQCT scans of patients with osteoarthritis or rheumatoid arthritis.

The ORMIR_XCT package contains four modules for processing HR-pQCT data of bones and joints: 1) automatic contouring of the periosteal surface, 2) JSW analysis, 3) BMD calculation, and 4) segmentation of trabecular bone. Analyses have been performed to compare outputs from the ORMIR_XCT package to results generated using IPL. Jupyter Notebook examples are provided to describe the workflows implemented in the ORMIR_XCT package.

# 3 – Comparison to IPL:

## Autocontour:

Automatic periosteal contouring (autocontour) was performed on a sample dataset of HR-pQCT images of the 2nd and 3rd distal interphalangeal (DIP) and trapeziometacarpal (TMC) joints (n = 59). The images came from 10 participants in an osteoarthritic group and 10 participants in a control group. This dataset is representative of typical data that is collected using HR-pQCT. Binarized images were obtained using both the IPL standard method for automatic periosteal contouring [4] and the ORMIR_XCT package. Segmentations were compared by computing DICE coefficients, Jaccard indices, and Hausdorff distances between images. The IPL method for automatic periosteal contouring was taken as the ground truth. Results of the segmentation comparison are shown in Table 1.

## Thickness:

The ORMIR_XCT package contains an open-source implementation of the algorithm for model-independent thickness estimation by Hildebrand *et al* [5]. This thickness estimation is used to compute several morphological measurements from HR-pQCT images, including trabecular thickness, trabecular separation, and cortical thickness. Given a binary mask describing a three-dimensional object as input, the algorithm first computes the distance map (via a distance transform), describing at each voxel the radius of the largest sphere centered at that voxel that is entirely within the object. Next, the local thickness map is computed by iterating through all voxels in the mask and assigning them a local thickness value that is the diameter of the largest sphere in the distance map that contains that voxel. The thickness of a structure can then be computed as the volume-weighted mean of the local thickness map, with optional adjustments for a minimum thickness value if desired to reduce the effect of surface noise.

Hildebrand *et al.* note an optimization for the algorithm where first, the distance ridge (also known as skeletonization or medial axis) is computed and only the distance map values on the distance ridge are used for local thickness map computation. We implemented this optimized version of the algorithm with the NumPy (v1.23.3) [6] and Numba (v0.56.3) [7] Python packages. Finally, we have noted and compensated for a discrepancy between the definition of the distance map by

Hildebrand *et al* and the outputs of many common open-source distance map algorithms (including those available in SimpleITK (v2.0.2) [8] and scikit-image (v0.19.2) [9]. Hildebrand *et al.* define their distance map as "the radius of the largest sphere centered at the point and still completely inside the structure". In contrast, most other distance transforms compute the distance from the center of a voxel inside the mask to the nearest voxel center outside the mask.

To compensate for this, we developed an oversampling distance transform function (Figure 1). In this function, an input mask is up-sampled to double the voxel size, and a sequence of morphological filters are applied to ensure that the newly introduced voxels on the boundary between the structure and the background are assigned as background voxels. A distance transform is applied to this mask, resulting in a distance map where the spheres fit entirely within the structure defined by the mask, and the resulting distance map is down-sampled to the original voxel size.

The user is given a choice, via a Boolean flag parameter in the *compute_local_thickness_from_mask* and *calc_structure_thickness_statistics* functions, to use the oversampling distance transform for thickness estimation. The choice is given because different users may have variable preferences in applying alternate definitions regarding the bounds of their structures (whether they be contained within the mask voxels or extend to the center of the neighbouring background voxels) and because using the normal (*i.e.,* not oversampled) distance transform will provide better congruence with thickness calculations produced using IPL.

To evaluate the thickness computation implemented in the ORMIR_XCT package, a set of synthetic shapes were generated to simulate bones of varying sizes. These shapes include solid and hollow spheres, solid and hollow cylinders, and plates of varying thicknesses [10]. The standard IPL joint space width (JSW) analysis workflow [11] was performed on the synthetic shape dataset using the following parameters: ridge_epsilon = 0.9, assign_epsilon = 1.8, suppress_boundary = 0, and version = 3. This set of parameters is currently used for JSW analysis of *in vivo* finger joints. ORMIR_XCT thickness was calculated using the oversampling distance transform function described above. Bland-Altman plots were generated to compare thickness results between algorithms (Figure 2).

**Bone Mineral Density:**

Images obtained from HR-pQCT scanners are saved using the manufacturer's proprietary file format (AIM). To process HR-pQCT images using Python, these files need to be converted to a file format that can be interpreted by common image processing Python libraries. The ITKIOScanco module is used in the ORMIR_XCT package to convert HR-pQCT images to other common medical image file types. However, the ITKIOScanco module converts HR-pQCT images to Hounsfield Units (HU) by default. Measures of BMD require the image to be calibrated in density units (mg of hydroxyapatite per cm$^3$). Several image unit conversions have been included in the ORMIR_XCT package to allow for conversion between HR-pQCT native units (termed Scanco units) and each of HU, density, and linear attenuation units. A sample dataset of HR-pQCT images of second and third metacarpal phalangeal (MCP) joints (n = 292) was used to compare BMD results between IPL and the ORMIR_XCT package. BMD was reported separately for the distal (DST) and proximal (PRX) segments of each joint (Figure 3).

**Trabecular Segmentation:**

A common operation performed in IPL is the application of a Gaussian smoothing filter followed by a global threshold-based image segmentation. The *gauss_seg* function in IPL has been reimplemented in the ORMIR_XCT package and the similarity of IPL and ORMIR_XCT trabecular segmentations were computed using DICE coefficients, Jaccard indices, and Hausdorff distances (Table 2). The same dataset used to test the automatic contouring module was used to compare IPL and ORMIR_XCT trabecular segmentation.

# References:

1. Brunet, S. C., Kuczynski, M. T., Bhatla, J. L., Lemay, S., Pauchard, Y., Salat, P., ... & Manske, S. L. (2020). The utility of multi-stack alignment and 3D longitudinal image registration to assess bone remodeling in rheumatoid arthritis patients from second generation HR-pQCT scans. BMC Medical Imaging, 20, 1-10.
2. Burt, L. A., Liang, Z., Sajobi, T. T., Hanley, D. A., & Boyd, S. K. (2016). Sex-and site-specific normative data curves for HR-pQCT. Journal of Bone and Mineral Research, 31(11), 2041-2047.
3. Brunet, S. C., Tse, J. J., Kuczynski, M. T., Engelke, K., Boyd, S. K., Barnabe, C., & Manske, S. L. (2021). Heterogenous bone response to biologic DMARD therapies in rheumatoid arthritis patients and their relationship to functional indices. Scandinavian Journal of Rheumatology, 50(6), 417-426.
4. Burghardt, A. J., Buie, H. R., Laib, A., Majumdar, S., & Boyd, S. K. (2010). Reproducibility of direct quantitative measures of cortical bone microarchitecture of the distal radius and tibia by HR-pQCT. Bone, 47(3), 519-528.
5. Hildebrand, T., & Rüegsegger, P. (1997). A new method for the model-independent assessment of thickness in three-dimensional images. Journal of microscopy, 185(1), 67-75.
6. Harris, C. R., Millman, K. J., Van Der Walt, S. J., Gommers, R., Virtanen, P., Cournapeau, D., ... & Oliphant, T. E. (2020). Array programming with NumPy. Nature, 585(7825), 357-362.
7. Lam, S. K., Pitrou, A., & Seibert, S. (2015, November). Numba: A llvm-based python jit compiler. In Proceedings of the Second Workshop on the LLVM Compiler Infrastructure in HPC (pp. 1-6).
8. Lowekamp, B. C., Chen, D. T., Ibáñez, L., & Blezek, D. (2013). The design of SimpleITK. Frontiers in neuroinformatics, 7, 45.
9. Van der Walt, S., Schönberger, J. L., Nunez-Iglesias, J., Boulogne, F., Warner, J. D., Yager, N., ... & Yu, T. (2014). scikit-image: image processing in Python. PeerJ, 2, e453.
10. Stok, K. S., Burghardt, A. J., Boutroy, S., Peters, M. P., Manske, S. L., Stadelmann, V., ... & SPECTRA Collaboration. (2020). Consensus approach for 3D joint space width of metacarpophalangeal joints of rheumatoid arthritis patients using high-resolution peripheral quantitative computed tomography. Quantitative imaging in medicine and surgery, 10(2), 314.
11. Kuczynski, M.T. (2023). Simulated Shapes for ORMIR_XCT Thickness Measurement Validation [Data set]. Zenodo.

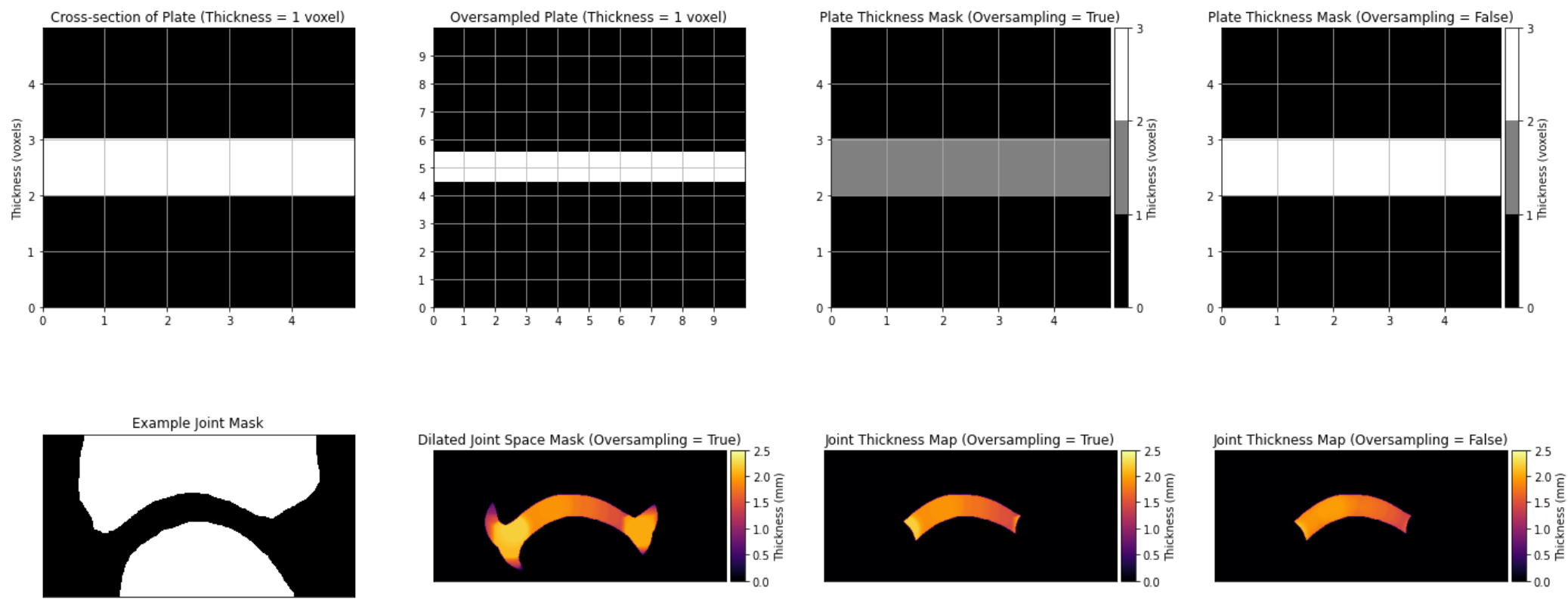

**Figure 1:** Examples of the thickness estimation algorithm implemented in the ORMIR_XCT package. Top row: thickness calculation on a 1 voxel thick plate demonstrating how thickness estimation is improved using the oversampling distance transform function (thickness = 1 voxel using oversampling, thickness = 2 voxels without oversampling). Bottom row: thickness calculation applied to a sample finger joint.

**Table 1:** DICE coefficients, Jaccard indices, and Hausdorff distances (voxels) between IPL and ORMIR_XCT automatic contouring implementations. All scans were obtained from a second-generation HR-pQCT scanner (XtremeCT2, Scanco Medical) with an isotropic voxel size of 61μm. DIP2: 2nd distal interphalangeal joint, DIP3: 3rd distal interphalangeal joint, TMC: trapeziometacarpal joint.

| Scan | DIP2 | | | | DIP3 | | | | TMC | | | |
|---|---|---|---|---|---|---|---|---|---|---|---|---|
| | DICE | Jaccard | Mean Hausdorff (voxels) | Maximum Hausdorff (voxels) | DICE | Jaccard | Mean Hausdorff (voxels) | Maximum Hausdorff (voxels) | DICE | Jaccard | Mean Hausdorff (voxels) | Maximum Hausdorff (voxels) |
| 1 | 0.97 | 0.94 | 0.05 | 10.68 | 0.98 | 0.96 | 0.03 | 9.06 | 0.98 | 0.96 | 0.06 | 21.02 |
| 2 | 0.96 | 0.92 | 0.06 | 7.81 | 0.94 | 0.89 | 0.15 | 18.68 | 0.98 | 0.97 | 0.04 | 19.03 |
| 3 | 0.97 | 0.94 | 0.05 | 8.12 | 0.97 | 0.93 | 0.07 | 17.32 | 0.98 | 0.96 | 0.06 | 25.67 |
| 4 | 0.96 | 0.93 | 0.06 | 10.05 | 0.98 | 0.96 | 0.03 | 9.11 | 0.95 | 0.91 | 0.26 | 37.36 |
| 5 | 0.97 | 0.94 | 0.06 | 10.49 | 0.97 | 0.94 | 0.05 | 12.41 | 0.98 | 0.96 | 0.07 | 22.67 |
| 6 | 0.96 | 0.93 | 0.07 | 12.37 | 0.96 | 0.92 | 0.08 | 14.73 | 0.99 | 0.97 | 0.03 | 20.61 |
| 7 | 0.98 | 0.97 | 0.02 | 6.40 | 0.99 | 0.98 | 0.01 | 3.61 | 0.98 | 0.96 | 0.08 | 26.32 |
| 8 | 0.97 | 0.93 | 0.06 | 20.62 | 0.98 | 0.96 | 0.04 | 13.34 | 0.94 | 0.89 | 0.51 | 45.29 |
| 9 | 0.98 | 0.97 | 0.02 | 9.70 | - | - | - | - | 0.98 | 0.95 | 0.06 | 21.21 |
| 10 | 0.98 | 0.96 | 0.03 | 8.60 | 0.97 | 0.95 | 0.04 | 18.00 | 0.96 | 0.91 | 0.23 | 33.73 |
| 11 | 0.98 | 0.96 | 0.03 | 5.48 | 0.99 | 0.97 | 0.02 | 8.06 | 0.99 | 0.97 | 0.03 | 22.14 |
| 12 | 0.97 | 0.94 | 0.05 | 9.00 | 0.96 | 0.92 | 0.09 | 27.29 | 0.97 | 0.95 | 0.09 | 22.00 |
| 13 | 0.98 | 0.96 | 0.02 | 9.00 | 0.98 | 0.96 | 0.03 | 8.60 | 0.97 | 0.95 | 0.09 | 21.66 |
| 14 | 0.98 | 0.96 | 0.03 | 6.00 | 0.99 | 0.97 | 0.02 | 7.28 | 0.99 | 0.97 | 0.03 | 17.92 |
| 15 | 0.98 | 0.95 | 0.04 | 14.87 | 0.98 | 0.97 | 0.02 | 7.68 | 0.99 | 0.97 | 0.02 | 18.11 |
| 16 | 0.96 | 0.93 | 0.07 | 10.68 | 0.99 | 0.97 | 0.02 | 7.00 | 0.99 | 0.97 | 0.03 | 18.14 |
| 17 | 0.99 | 0.98 | 0.02 | 6.16 | 0.98 | 0.97 | 0.03 | 9.43 | 0.99 | 0.98 | 0.03 | 23.26 |
| 18 | 0.99 | 0.98 | 0.01 | 4.47 | 0.99 | 0.98 | 0.01 | 4.90 | 0.99 | 0.97 | 0.03 | 15.30 |
| 19 | 0.98 | 0.97 | 0.02 | 4.24 | 0.99 | 0.97 | 0.02 | 23.60 | 0.99 | 0.97 | 0.03 | 22.20 |
| 20 | 0.98 | 0.96 | 0.04 | 11.75 | 0.99 | 0.97 | 0.02 | 4.36 | 0.99 | 0.97 | 0.04 | 24.45 |

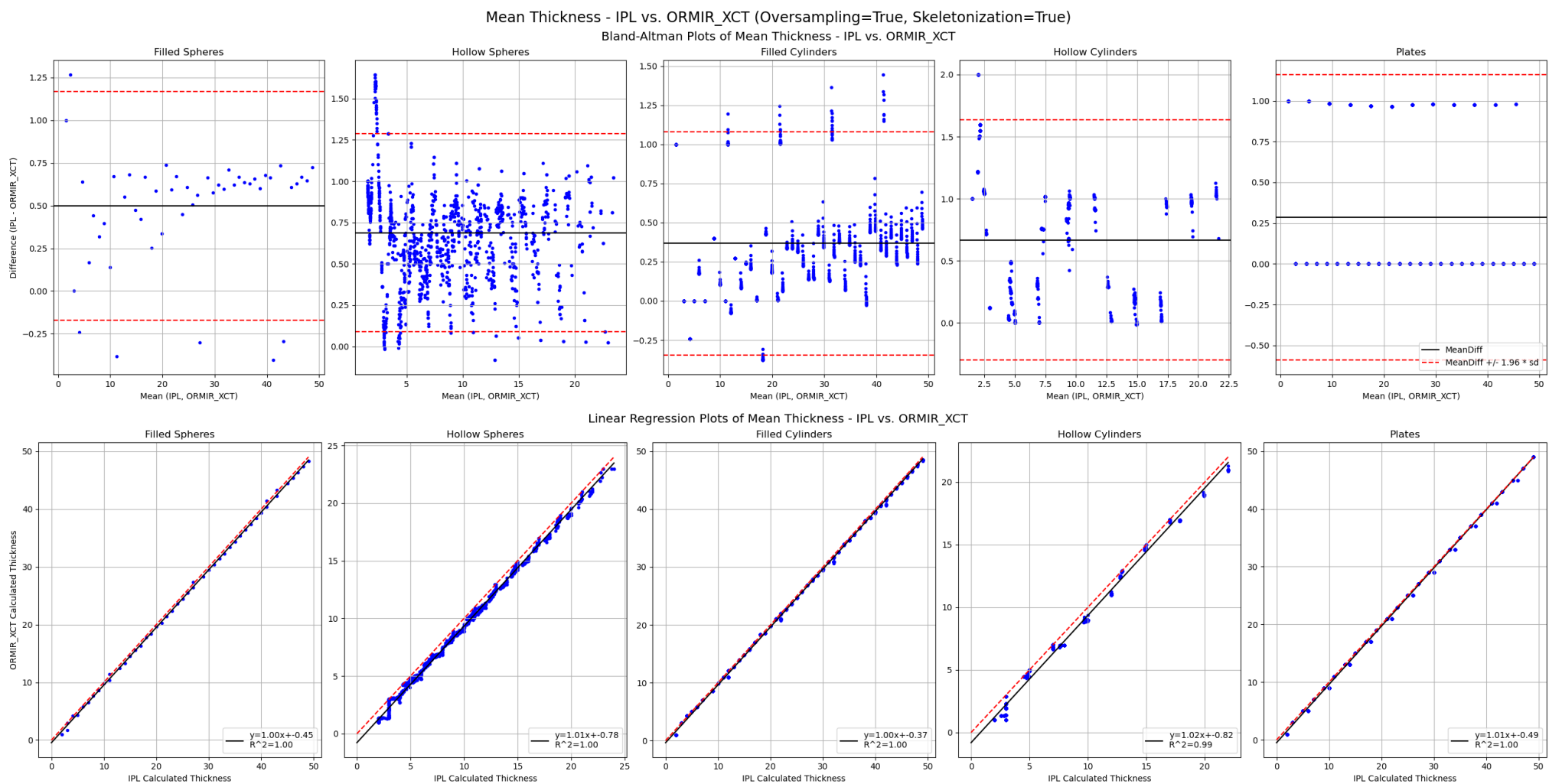

**Figure 2:** Bland-Altman and regression plots of mean joint space thickness computed using IPL and the ORMIR_XCT package. Top row: black lines indicate mean difference and red dotted line shows the limits of agreement. Bottom row: solid black line indicates the linear regression between IPL and ORMIR_XCT computed thickness and red dotted lines indicate a line of unity.

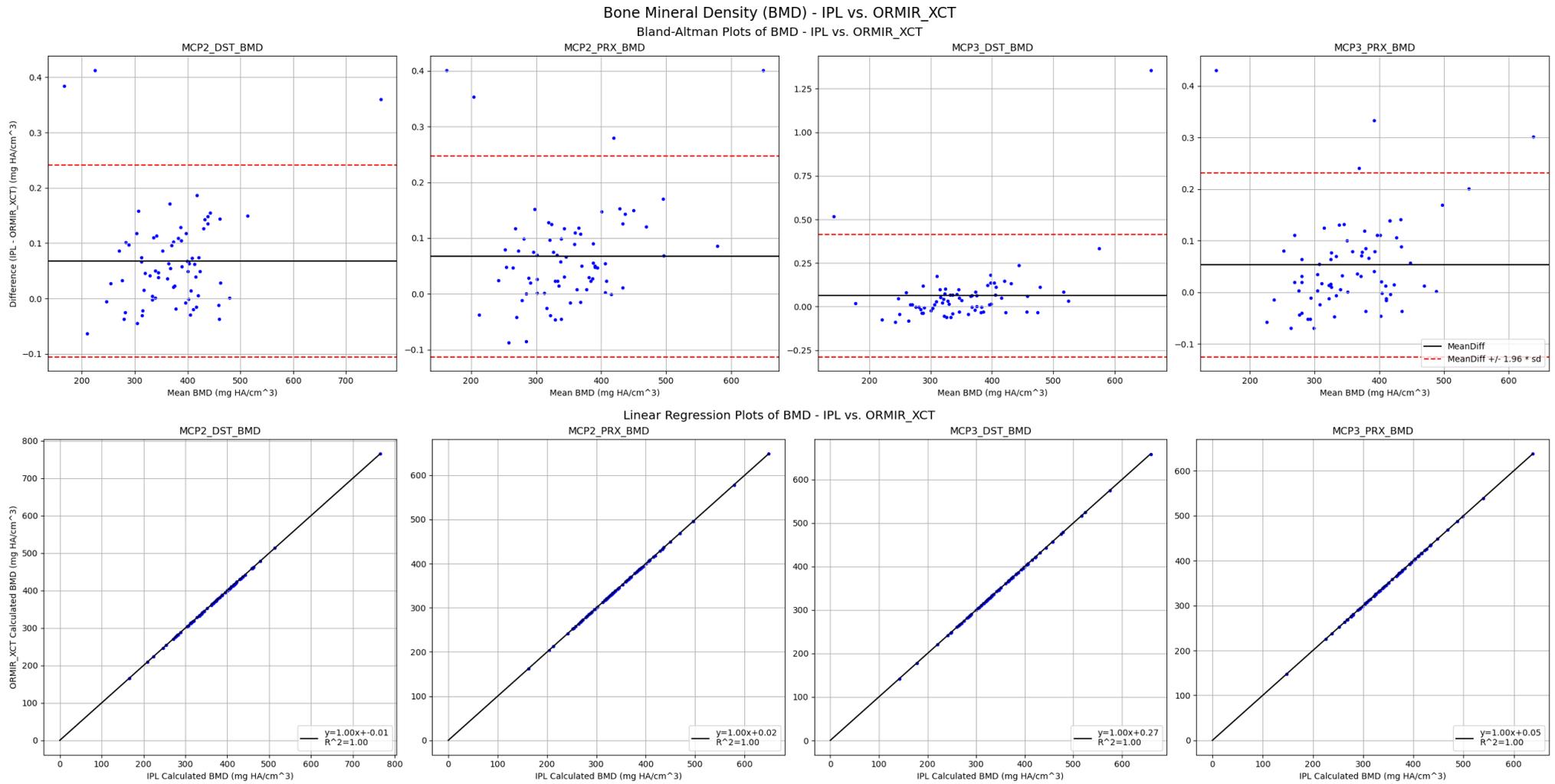

**Figure 3:** Bone mineral density (mg HA/cm$^3$) linear regression and Bland-Altman plots comparing IPL and the ORMIR_XCT computed bone mineral density (BMD). All scans were obtained from a second-generation HR-pQCT scanner (XtremeCT2, Scanco Medical) with an isotropic voxel size of 61μm. MCP2: 2$^{nd}$ metacarpal phalangeal joint, MCP3: 3$^{rd}$ metacarpal phalangeal joint, DST: distal, PRX: proximal.

**Table 2:** DICE scores, Jaccard indices, and Hausdorff distances (voxels) for comparison of trabecular segmentation performed using IPL versus the ORMIR_XCT package. All scans were obtained from a second-generation HR-pQCT scanner (XtremeCT2, Scanco Medical) with an isotropic voxel size of 61μm. For each algorithm, Gaussian smoothing was performed with sigma = 0.5. Subsequently, segmentation was performed with lower and upper thresholds of 1,170 and 10,000 HU, respectively. DIP2: second distal interphalangeal joint, DIP3: third distal interphalangeal joint, TMC: trapeziometacarpal joint.

| Scan | DIP2 | | | | DIP3 | | | | TMC | | | |
|---|---|---|---|---|---|---|---|---|---|---|---|---|
| | DICE | Jaccard | Mean Hausdorff (voxels) | Maximum Hausdorff (voxels) | DICE | Jaccard | Mean Hausdorff (voxels) | Maximum Hausdorff (voxels) | DICE | Jaccard | Mean Hausdorff (voxels) | Maximum Hausdorff (voxels) |
| 1 | 0.99 | 0.98 | 0.01 | 16.55 | 0.99 | 0.97 | 0.02 | 16.25 | 0.99 | 0.98 | 0.02 | 40.16 |
| 2 | 0.99 | 0.98 | 0.02 | 10.10 | 0.99 | 0.97 | 0.02 | 11.36 | 0.99 | 0.98 | 0.01 | 19.34 |
| 3 | 0.99 | 0.98 | 0.01 | 9.22 | 0.99 | 0.98 | 0.01 | 12.21 | 0.99 | 0.98 | 0.02 | 37.66 |
| 4 | 0.99 | 0.98 | 0.01 | 13.49 | 0.99 | 0.98 | 0.01 | 11.36 | 0.99 | 0.97 | 0.03 | 36.46 |
| 5 | 0.99 | 0.98 | 0.01 | 22.11 | 0.99 | 0.98 | 0.01 | 18.00 | 0.99 | 0.98 | 0.02 | 32.40 |
| 6 | 0.99 | 0.97 | 0.02 | 13.19 | 0.99 | 0.98 | 0.02 | 16.31 | 0.99 | 0.98 | 0.02 | 29.00 |
| 7 | 0.99 | 0.99 | 0.01 | 8.54 | 0.99 | 0.99 | 0.01 | 8.06 | 1.00 | 0.99 | 0.01 | 23.22 |
| 8 | 1.00 | 0.99 | 0.01 | 10.30 | 0.99 | 0.99 | 0.01 | 15.03 | 1.00 | 0.99 | 0.01 | 33.62 |
| 9 | 0.99 | 0.99 | 0.01 | 5.83 | - | - | - | - | 1.00 | 0.99 | 0.01 | 35.18 |
| 10 | 0.99 | 0.99 | 0.01 | 10.63 | 0.99 | 0.99 | 0.01 | 9.43 | 1.00 | 0.99 | 0.01 | 30.50 |
| 11 | 0.99 | 0.99 | 0.01 | 9.00 | 0.99 | 0.99 | 0.01 | 9.95 | 1.00 | 0.99 | 0.01 | 31.95 |
| 12 | 0.99 | 0.98 | 0.01 | 9.49 | 1.00 | 0.99 | 0.01 | 15.59 | 0.99 | 0.98 | 0.01 | 21.28 |
| 13 | 0.99 | 0.99 | 0.01 | 13.75 | 0.99 | 0.99 | 0.01 | 13.15 | 0.99 | 0.98 | 0.05 | 36.94 |
| 14 | 0.99 | 0.98 | 0.01 | 9.64 | 0.99 | 0.99 | 0.01 | 11.66 | 1.00 | 0.99 | 0.02 | 18.38 |
| 15 | 0.99 | 0.99 | 0.01 | 13.19 | 0.99 | 0.99 | 0.01 | 12.73 | 0.99 | 0.99 | 0.03 | 33.20 |
| 16 | 0.99 | 0.99 | 0.01 | 17.49 | 0.99 | 0.99 | 0.01 | 11.05 | 1.00 | 0.99 | 0.01 | 23.15 |
| 17 | 0.99 | 0.99 | 0.01 | 5.92 | 0.99 | 0.99 | 0.01 | 6.63 | 1.00 | 1.00 | 0.00 | 27.22 |
| 18 | 0.99 | 0.99 | 0.01 | 8.37 | 0.99 | 0.99 | 0.01 | 9.95 | 1.00 | 0.99 | 0.01 | 24.92 |
| 19 | 0.99 | 0.99 | 0.01 | 12.33 | 0.99 | 0.99 | 0.01 | 12.08 | 1.00 | 0.99 | 0.02 | 29.68 |
| 20 | 0.99 | 0.99 | 0.01 | 8.06 | 0.99 | 0.99 | 0.01 | 7.87 | 1.00 | 0.99 | 0.01 | 25.81 |